# Direct Measurement of Biexcitons in Monolayer WS$_2$

M.A. Conway[1,2], J.B Muir[1,2], S.K. Earl[1,2], M. Wurdack[3], R. Mishra[1,2], J.O. Tollerud[1,2] & J.A. Davis[1,2]

[1]*Optical Sciences Centre, Swinburne University of Technology, Victoria 3122, Australia*

[2]*ARC Centre of Excellence in Future Low-Energy Electronics Technologies, Swinburne University of Technology, Victoria, 3122, Australia*

[3]*ARC Centre of Excellence in Future Low-Energy Electronics Technologies and Nonlinear Physics Centre, Research School of Physics, The Australian National University, Canberra, ACT, 2601, Australia*

*jdavis@swin.edu.au*

**Abstract**

The optical properties of atomically thin transition metal dichalcogenides (TMDCs) are dominated by Coulomb bound quasi-particles, such as excitons, trions, and biexcitons. Due to the number and density of possible states, attributing different spectral peaks to the specific origin can be difficult. In particular, there has been much conjecture around the presence, binding energy and/or nature of biexcitons in these materials. In this work, we remove any ambiguity in identifying and separating the optically excited biexciton in monolayer WS$_2$ using two-quantum multidimensional coherent spectroscopy (2Q-MDCS), a technique that directly and selectively probes doubly-excited states, such as biexcitons. The energy difference between the unbound two-exciton state and the biexciton is the fundamental definition of biexciton binding energy and is measured to be 26 ± 2 meV. Furthermore, resolving the biexciton peaks in 2Q-MDCS allows us to identify that the biexciton observed here is composed of two bright excitons in opposite valleys.

**Introduction**

Coulomb correlations between photoexcited electrons and holes in atomically thin transition metal dichalcogenides (TMDCs) mediate a variety of bound electronic excitations, such as excitons [1-3], trions [4, 5], and biexcitons [1, 4, 6]. The binding energy of these quasi-particles is greatly enhanced over those in 3D material systems because the screening of the Coulomb interaction is greatly reduced [7]. The nature and binding energy of biexcitons in these materials has, however, been a topic of debate, with reported binding energies ranging from more than 60 meV [8-11] to as small as 20meV [6, 12-16], while others have suggested that optically-induced biexciton formation is not evident in time resolved measurements [17].

Many of the early experimental observations of biexcitons in monolayer TMDCs relied on photoluminescence spectroscopy to identify peaks below the exciton energy, arising from radiative relaxation from the biexciton to the exciton state [2, 4, 6, 8-12, 15, 16]. These have been combined with intensity dependent measurements to show the quadratic power dependence expected for biexcitons. There are, however, many different peaks at similar energies, including those associated with trions and dark excitons, which should grow linearly with pump fluence, as well as charged biexcitons [12, 15], defect-bound excitons [18], phonon sidebands [19] and electron plasma excitations [20], which, like biexcitons, grow super-linearly with pump fluence. The ability to resolve all these peaks can be further complicated by static disorder, which leads to inhomogeneous broadening of the spectral peaks. However, recent efforts in encapsulating the monolayers in hexagonal boron nitride has mitigated these effects and led to narrow, resolvable peaks, which has aided the identification of biexcitons [12, 21].

In tungsten based semiconducting TMDCs, such as $WS_2$ and $WSe_2$ the situation is further complicated by the band ordering. Specifically, the spin-orbit splitting of the conduction band results in the lowest energy exciton being a dark, spin-forbidden exciton, with the bright exciton transition 10-30 meV higher in energy [22]. It then becomes important to consider the timescale of the measurements and the excitation energy, as optically excited bright excitons or free-carriers can relax into these dark exciton states, which also support optically allowed transitions to a biexciton state consisting of one bright and one dark exciton.

Photoluminescence measurements on encapsulated $WSe_2$ have attributed a peak located 18 meV below the A-exciton peak to the emission from biexciton to exciton, specifically, a biexciton involving a dark plus a bright exciton in opposite valleys [12]. They also identify a charged biexciton located 49meV below the A-exciton emission, consistent with theory predictions [14]. This may be the origin for earlier measurements reporting a biexciton binding energy of around 40-60 meV [8-11].

In time-resolved measurements an induced absorption peak is seen on the low energy side of the exciton peak and has been attributed to absorption from a single exciton state to a biexciton state [6, 8]. Polarization resolved measurements show this peak to be present only in the case of cross-circular polarization, as would be expected for the biexciton transition, which must involve excitons of different spin. Other reports, however, attribute this induced absorption peak to band gap renormalization [17, 23, 24].

Multi-dimensional coherent spectroscopy (MDCS), which reveals correlations of absorption and emission energy, provides a clearer picture, and has been used to identify peaks corresponding to transitions from exciton to biexciton and charged biexciton states in monolayer $WSe_2$ [15]. In contrast, MDCS measurements on monolayer $WS_2$ saw no signatures of biexcitons [25], while in $MoS_2$ monolayers the interpretation of MDCS measurements led to conclusions that the photoinduced absorption observed is due to band gap renormalization and that there is no direct excitation of the biexciton [17].

Here, we remove any ambiguity by using two-quantum MDCS (2Q-MDCS), which selectively excites and probes doubly excited states such as biexcitons, eliminating any contributions from singly excited states, including band-gap renormalization. 2Q-MDCS has been used previously to identify and quantify the properties of biexcitons in semiconductor quantum-wells [26-28]. In this type of experiment, the energy of the biexciton ($XX^b$) and the unbound two-exciton state (XX) can be measured directly, as indicated in Fig. 1 (b). The energy difference between these states is the fundamental definition of the biexciton binding energy ($E_b$). In contrast, all the other approaches mentioned above obtain $E_b$ indirectly, by comparing the energy of the $XX^b$ – X transition to that of the X – g transition [27]. In the 2Q-MDCS measurements reported here we are able to identify the two-photon-optically-excited biexciton in monolayer $WS_2$ as consisting of two bright excitons from different valleys, with a binding energy of 26 ± 2 meV.

**Experimental Methods**

Third-order MDCS is based on a transient four-wave mixing (FWM) experiment with the phase and amplitude of the signal measured as a function of the emission energy and the delays between the three excitation pulses [28, 29]. We use a box-CARS geometry, as depicted in Figure 1a, with each of the three incident pulses on the corners of the box and labelled $k_1$, $k_2$, $k_3$. The FWM signal is emitted in background-free directions given by conservation of momentum. We measure the signal emitted in the direction given by -$k_1$ + $k_2$ + $k_3$, which is the fourth corner of the box, and which is overlapped with a fourth and much weaker beam, referred to as the local oscillator (LO). The LO interferes with the signal in the spectrometer to give a spectral interferogram, from which we can determine both the amplitude and phase of the signal. Phase stability between each of the excitation pulses and the LO is essential and is maintained passively by ensuring all beams are incident on common optics [30, 31].

The most common type of 2D MDCS spectra is referred to as a 1Q rephasing spectrum and is acquired when the $k_1$-pulse arrives first. After the first pulse, the system is in a coherent superposition of states separated by the optical photon energy (a 1Q-coherence). The second pulse, $k_2$, converts the 1Q-coherence into a population (or low energy coherence) in either the excited or ground electronic state. The third pulse then converts this into a third-order 1Q-coherence, and corresponding macroscopic polarization, which radiates the signal. The measured spectral interferogram of the signal gives the amplitude and phase of the signal as a function of the emission energy, $E_3$. As the delay between the first two pulses, $t_1$, is varied, the phase of the coherent superposition excited by $k_1$ continues to evolve, which is mapped onto the phase of the signal. The spectrally resolved signal is then Fourier transformed with respect to $t_1$ to give the 1Q rephasing 2D spectrum, which correlates the energy of the coherence excited by the first pulse, $E_1$ (often approximated as the absorption energy) with the emission energy ($E_3$). In the case of biexcitons, the third pulse $k_3$ drives the system into a third-order 1Q-coherence between the exciton and biexciton states, which radiates as the signal. In 1Q rephasing 2D spectra, biexcitons therefore appear as a peak with excitation energy $E_1$ equal to the exciton transition energy but at an emission energy $E_3$ below the exciton energy, by an amount given by the biexciton binding energy. Because of the polarization selection rules shown in Figure 1d, and the Pauli exclusion principle, biexcitons are typically observed with cross-circularly polarized pulse sequences [8] or linearly polarized pulses, but are not observed when all pulses are co-circularly polarized. In TMDCs, this requires an exciton in each valley as has been identified previously [8-10, 12-16, 32, 33].

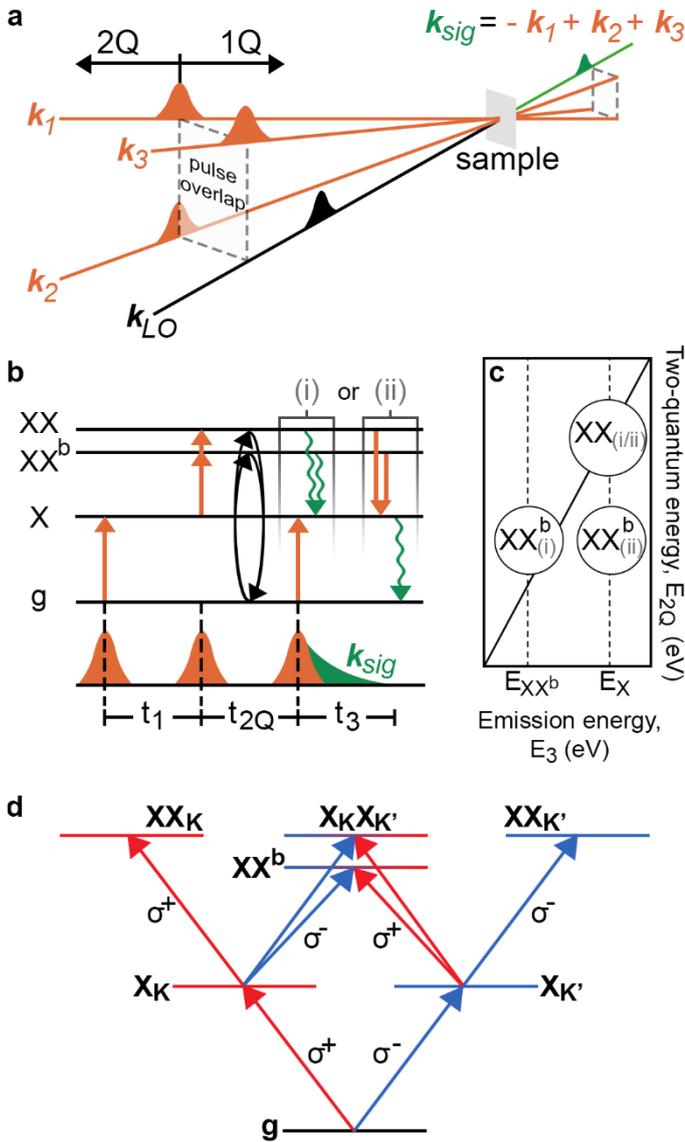

Figure 1: **a** Four beam box geometry utilized in our experiments. In our setup, the pulse $k_1$ is scanned in time and when it arrives before the other pulses the signal generated is a 1Q rephasing signal. When $k_1$ arrives last, the system is driven into a 2Q coherence by $k_2$ and $k_3$, indicated by the black arrows in **b**. This 2Q coherence is reduced to a 1Q coherence upon interaction with the third pulse after some time $t_{2Q}$ which radiates as the signal. There are two possible radiative signal pathways, (i) and (ii), which are energetically degenerate for correlated excitons (XX), but differ by the binding energy for biexcitons ($XX^b$). **c** Schematic of the expected 2Q spectral peaks from XX and $XX^b$ via the different pathways (i) and (ii) indicated in **b**. **d** Energy level diagram of monolayer TMDs showing pathway to unbound (XX) and bound ($XX^b$) two exciton states.

Changing the pulse ordering so that the $k_1$ pulse arrives last enables the acquisition of 2Q-MDCS. In this case, the first pulse still generates a 1Q coherence between the ground state and the exciton state, but the second pulse drives the transition from the singly excited state to the doubly excited state, generating a coherent superposition between the ground and doubly excited state, which we refer to as a 2Q coherence, as depicted in Figure 1b. The phase evolution of this 2Q coherence over the time interval between the second and third pulses, $t_{2Q}$, is mapped on to the signal phase, and the subsequent Fourier transform of the signal with respect to $t_{2Q}$ gives the energy of the doubly excited state(s). This is typically plotted on the vertical axis in the associated 2Q-MDCS spectra. The third pulse reduces the 2Q coherence back to a 1Q coherence via one of the two possible pathways, which leads to the emitted signals, as depicted in Figure 1b: the emission in (i) is from a doubly excited state to a singly excited state while in (ii) it is from the singly excited state to the ground state. In the case of a two-exciton state with no interactions between the pair of excitons these two different pathways will cancel, leading to no signal. Where there are excitonic interactions, either pairwise or through many body effects, an asymmetry between the ground to single-exciton transition compared to the single-exciton to the two-exciton transition emerges, leading to imperfect cancellation and a measurable signal [26, 27].

For biexcitons, there is an obvious asymmetry in the transition energy, which also leads to peaks from the biexciton appearing at $E_{2Q}$ energy lower than twice the exciton energy, as indicated in Figure 1c, by the peaks labelled XX[b]. There are two peaks at different $E_3$ values arising from the different energies associated with the two different signal emission pathways indicated in Figure 1b. In addition to the biexciton, another peak may appear on the diagonal line corresponding to $E_{2Q} = 2E_X$, labelled XX, which originates from a correlated two-exciton state. This is an unbound two-exciton state where many-body effects introduce an asymmetry that allows peaks in the 2Q-MDCS spectra [26, 34].

We performed these 1Q- and 2Q-MDCS measurements on monolayer $WS_2$ exfoliated from a bulk $WS_2$ crystal and transferred to a $SiO_2$/Si substrate, as detailed in the supplementary information (SI). The laser spectrum was centered near 2 eV (~620 nm) with FWHM of 26nm (Figure S2), so that it covered the exciton energy and the energy of all other exciton complexes. Fluences were kept below 2 $\mu Jcm^{-2}$ per pulse to ensure the contributions of signals beyond the $\chi^{(3)}$ regime are insignificant.

**Results and Discussion**

The 1Q rephasing 2D spectra acquired using cross- and co-circularly polarized pulse sequences are shown in Figure 2a and b, respectively. The cross-circular spectrum is acquired by a series of excitation pulses with alternating helicity and features 2 peaks: a diagonal peak (a peak for which $E_1=E_3$) corresponding to excitation and emission from the exciton (X), and a cross peak redshifted from the exciton emission energy which we attribute to the neutral biexciton (XX[b]). In contrast, the co-circular spectrum in Figure 2b is acquired by a series of excitation pulses with the same helicity and shows only the X peak on the diagonal. The disappearance of the biexciton peak in a co-circular polarized excitation sequence is consistent with the optical selection rules, wherein biexcitons must consist of two excitons with opposite spin. Biexcitons made up of two excitons with the same spin is forbidden due to Pauli blocking [26, 27, 34, 35]. From the energy difference between the XX[b] and X peaks we obtain a biexciton binding energy of 24±4 meV. This spectrum is qualitatively comparable to prior 1Q MDCS measurements in $MoSe_2$ [15], however, there remains some conjecture around the origin of such peaks in the MDCS spectra of $WS_2$, since similar peaks can arise in the case of band gap renormalization [17].

The 2Q-MDCS spectra in Figure 2c and d remove any ambiguity in identifying and separating the biexciton. In both cross- and co-circular 2Q spectra there is a peak on the diagonal corresponding to unbound correlated excitons (XX) at $E_{2Q}$ equal to twice the single exciton energy (4.18 eV), and emission energy, $E_3$, equal to the exciton energy (2.09 eV). In the case of co-circularly polarized pulses, this is the only peak and there is no biexciton peak, as expected, due to Pauli blocking.

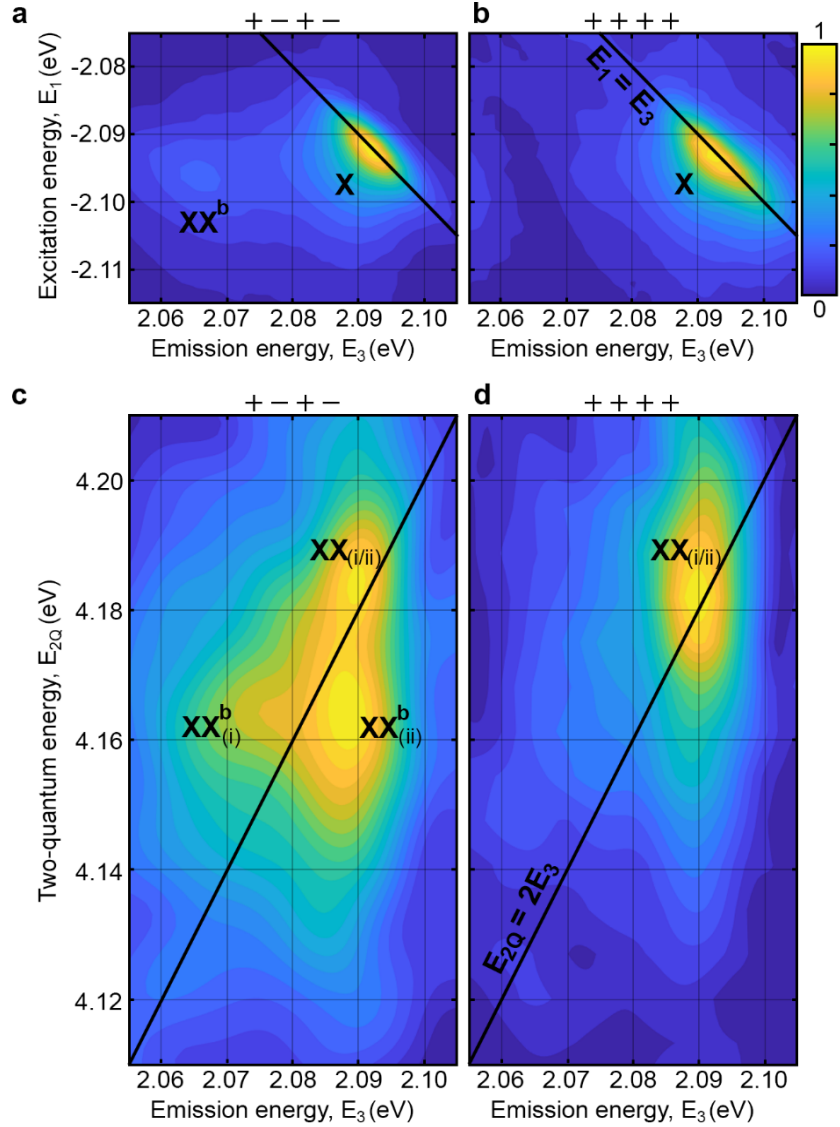

*Figure 2:* **a** *1Q cross-circular amplitude spectrum showing a bright exciton peak (X) and the biexciton ($XX^b$) redshifted in emission energy ($E_3$) by the biexciton binding energy.* **b** *The 1Q co-circular amplitude spectrum features an exciton peak but lacks the biexciton peak because of the optical selection rules associated with the bandstructure of monolayer $WS_2$.* **c** *2Q cross-circular amplitude spectrum shows a correlated exciton (XX) peak at twice the exciton energy in $E_{2Q}$ and two dominant biexciton ($XX^b$) peaks below XX by the biexciton binding energy. The two biexciton peaks are separated by the biexciton binding energy in $E_3$ and arise due to the two possible interactions with the third pulse as depicted in Figure 1b.* **d** *2Q co-circular amplitude spectrum lacks the biexciton peak, consistent with the optical selection rules of monolayer TMDCs. Contours are plotted at 5% intervals.*

For the cross-circularly polarized pulse sequence (Figure 2c), biexcitonic 2Q coherences can be excited (see SI for details of the pathways), leading to two peaks at $E_{2Q}$ = 4.163 eV, 26 meV below the correlated two-exciton peak. The different $E_3$ values for these peaks arise from the two signal pathways shown in Figure 1b. The higher energy peak at $E_3 \sim 2.09$ eV, corresponds to the single exciton to the ground state transition and arises from pathway (ii) in Figure 1b. The lower energy peak comes from pathway (i) in Figure 1b and corresponds to the biexciton to single exciton transition, with the emission energy $E_3$ corresponding to the exciton energy minus the biexciton binding energy.

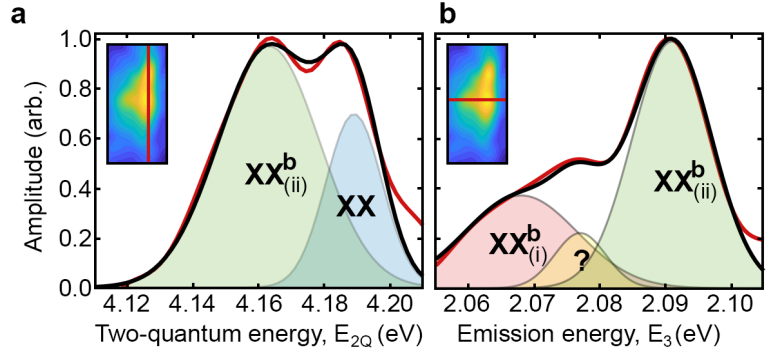

*Figure 3: **a** Vertical slice through Figure 2c (red) fit with two Gaussian functions (black). The energy difference between correlated (blue) and bound (green) two-exciton state peaks yields the biexciton binding energy 26±2 meV. **b** Horizontal slice through Figure 2c (red) fit with three Gaussian functions (black). The energy difference between the red and green biexciton peaks (24±4 meV) also corresponds to the biexciton binding energy. However, there is increased uncertainty in using this difference in emission energy for the same potential to overlay this signal contribution with other effects, similar to the issues faced by photoluminescence and pump-probe measurements. Uncertainties derived from the 95% confidence bounds of the fits.*

The energy difference between the correlated, unbound two-exciton state (XX) and the bound biexciton state ($XX^b$) is the fundamental definition of biexciton binding energy. Because 2Q-MDCS directly measures the two-exciton state energies, the biexciton binding energy can be determined reliably and directly. To do this, a vertical slice through the cross-circular 2Q 2D spectrum in Figure 2c, taken at the exciton emission energy in $E_3$, is plotted in red in Figure 3a. The slice is fit with the sum of two Gaussians (black line), with the individual Gaussians indicated by the shaded regions corresponding to the XX and $XX^b$ peaks. From this fit, and the splitting between the two Gaussians, we extract a biexciton binding energy in monolayer $WS_2$ of 26 ± 2 meV, where the uncertainty comes primarily from the width of the peaks. This value is consistent with theory calculations [14] which determine a binding energy of 23.9 ± 0.5 meV for $WS_2$, larger than for the other semiconducting TMDC monolayers.

In order to directly compare these measurements with photoluminescence, transient absorption and 1Q MDCS [15], where the biexciton binding energy is determined as the energy difference between the $XX^b \leftrightarrow X$ transition and the $X \leftrightarrow g$ transition, we consider the $E_3$ energy difference between $XX^b_{(i)}$ and $XX^b_{(ii)}$. To quantify these energies, we take a horizontal slice through the two $XX^b$ peaks in Figure 2c at $E_{2Q}$ = 4.163 eV (red line in Figure 3b). Three Gaussians are required to fit this slice, with peaks at 2.067, 2.077, and 2.091 eV. Taking the highest energy peak as arising from the $X \rightarrow g$ emission ($XX^b_{(ii)}$) and the lowest energy peak as the $XX^b \rightarrow X$ emission ($XX^b_{(i)}$), gives a biexciton binding energy of 24 ± 4 meV. This is consistent with the binding energy obtained from the $E_{2Q}$ energies, and somewhat higher than values measured in transient absorption [6] and PL [12] for $WSe_2$ and 1Q-MDCS for $MoSe_2$ [15], which is also consistent with theory predictions [14]. Obtaining the biexciton binding energy in this manner can, however, give inaccurate values in some circumstances [27]. This arises largely because the $XX^b \leftrightarrow X$ transition depends on the overlap of the biexciton and exciton wavefunctions which can weight subsets of the biexciton and exciton states that may not reflect the precise energy of either relative to the ground state [27]. This weighting of different components may also explain the additional peak between the exciton and biexciton emission peaks in Fig. 3b. This additional peak is ~9 meV higher in energy than the $XX^b_{(i)}$ peak, which is consistent with the splitting observed in the biexciton fine structure in $WSe_2$ [6]. In principle, signatures of the biexciton fine structure should also be separated in $E_{2Q}$, however, the spectral broadening of the peaks along the $E_{2Q}$ axis prevents us from being able to resolve whether there are any other distinct peaks. In these measurements the broadening arises from a combination of inhomogeneous broadening and excitonic interactions [36]. Future experiments with WS$_2$ monolayers encapsulated in hexagonal boron nitride may reduce these line widths and better resolve the fine-structure of the biexciton.

In addition to providing a more reliable means to confirm the direct excitation of biexcitons and quantify the biexciton binding energy, resolving the XX$^b$ peaks in 2Q-MDCS allows us to identify the nature of the biexciton. The biexciton observed here is composed of two bright excitons from opposite valleys: a bright-bright intervalley biexciton. If bright-dark biexcitons were the dominant exciton complex then we would expect to observe XX$^b$ signals at E$_{2Q}$ values given by twice the bright exciton energy minus the conduction band splitting (~26-30 meV [37, 38]) minus the biexciton binding energy (~23-26 meV); which would place it more than 50 meV below the XX peak. Despite the spectral range of our measurements extending to these lower energies, no such peak was observed. Bright-dark biexcitons have been indicated in photoluminescence experiments which measure the time integrated response of the sample [12]. At the longer timescales probed in photoluminescence, the dark exciton makes up the bulk of the excitonic population due to its favorable energy relative to the bright exciton. In contrast, MDCS measurements only directly excite the bright excitons and at the fs timescales probed, minimal relaxation to the dark exciton is expected. Furthermore, the relaxation pathways available are expected to include incoherent processes, so bright-dark biexcitons, even if they are present, are not expected to contribute to these coherent measurements except perhaps as an additional decoherence channel.

**Conclusion**

Using 2Q MDCS we have directly observed biexcitons in monolayer WS$_2$ by spectrally separating them from the correlated unbound two-exciton state. The results presented here extend upon previous MDCS measurements of monolayer TMDCs and supports attributing signals below the exciton energy to biexciton formation instead of bandgap renormalization for pulse fluences below 2 μJcm$^{-2}$. In addition, we experimentally resolve the biexciton binding energy in monolayer WS$_2$, (i.e. the energy difference between correlated (XX) and bound (XX$^b$) two-exciton states) to be 26 ± 2 meV from fits to slices of our 2Q spectrum. This value is below the intrinsic conduction band splitting in monolayer WS$_2$, confirming that the biexcitons observed here, on fs timescales, are bright-bright intervalley biexcitons.

**Acknowledgements**

This work was funded by the Australian Research Council Centre of Excellence in Future Low-Energy Electronics Technology (CE170100039).

## Supplementary Information

<u>Further Experimental Details</u>

The monolayer $WS_2$ used in these experiments was exfoliated from a bulk $WS_2$ crystal from HQ Graphene. It was transferred onto a silicon substrate with a 285 nm thick $SiO_2$ layer. An optical microscope image of the flake is shown in Figure S1. The laser spot diameter (FWHM) for the MDCS experiments was 45 µm, covering the full width of the sample across the middle region vertically.

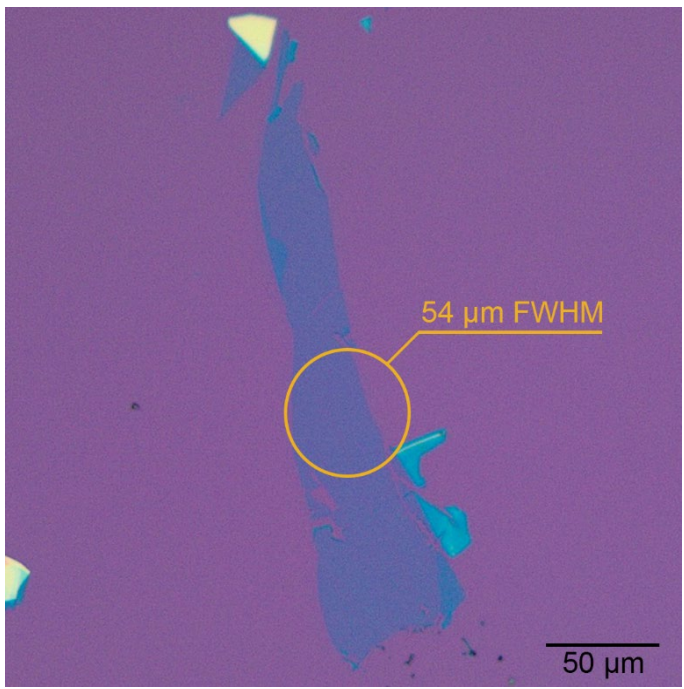

*Figure S1: Exfoliated monolayer $WS_2$ used in experiments. The scale bar is 50 µm, and the sample is approx. 225x40 µm. The 54 µm FWHM of a pulse is depicted by the orange circle, and indicates the location on the sample that measurements were conducted.*

The laser pulses used for the MDCS experiments were generated in a non-collinear optical parametric amplifier (Light Conversion Orpheus-N) pumped by the third harmonic of the 1030 nm, 220 fs pulses from a Yb:KGW laser amplifier (Light Conversion Pharos) at a repetition rate of 10.4 kHz. The pulse spectrum is shown in Fig. S2, centred at 620nm, and with a FWHM of 26.4 nm covers the exciton energy and the energy of all other exciton complexes at temperatures of 4K. The pulses were compressed using a multiphoton intrapulse interference phase scan (MIIPS) [39] with the spectral phase correction applied by the pulse shaper. The compressed pulse durations for all pulses were measured to be between 22 and 24 fs. The pulse fluences were kept below 2 $\mu Jcm^{-2}$ to ensure minimal excitation induced dephasing. Fluence dependent measurements were conducted to ensure this was the case and that with this fluence the dominant signal comes from the expected third-order response.

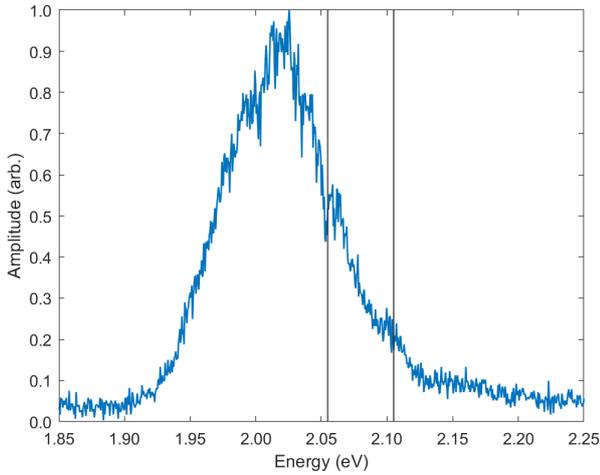

*Figure S2: Local oscillator spectrum. The vertical lines indicate the energy region windowed in our 2Q-MDCS figures of the main text.*

Full pathways for polarization resolved 2Q MDCS measurements:

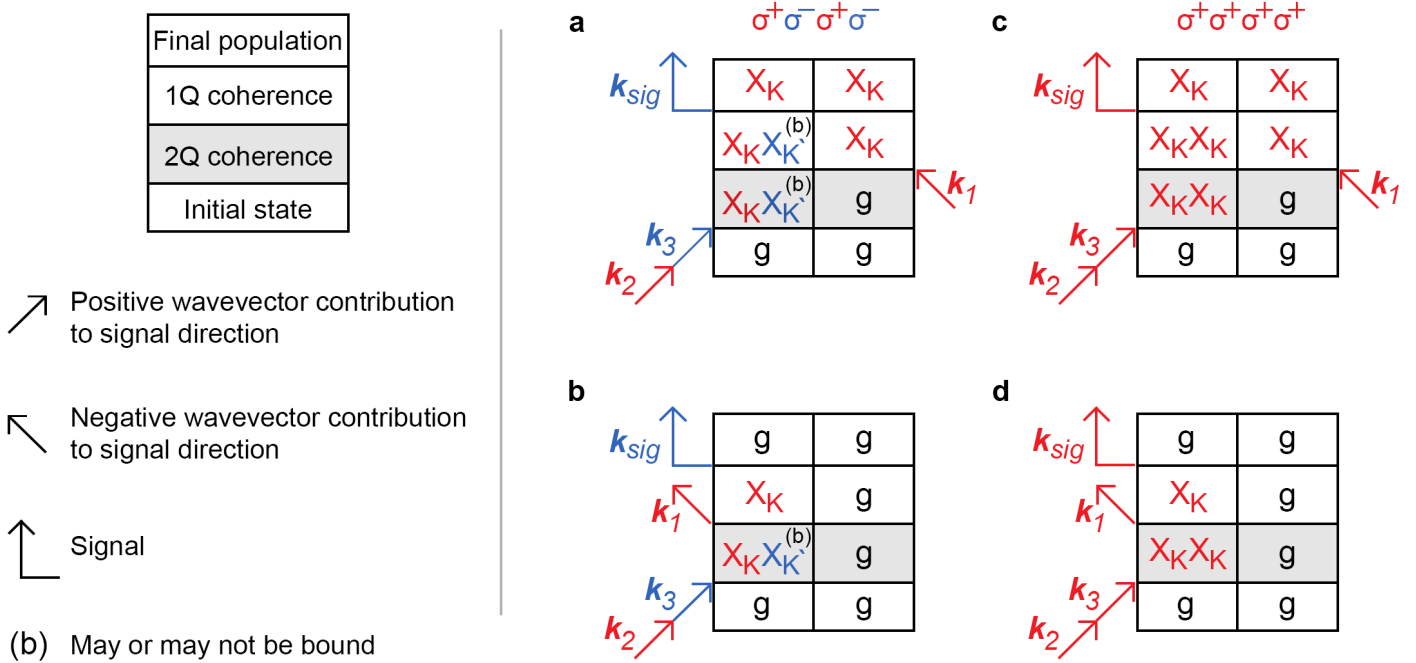

*Figure S3: The pathways for 2Q-MDCS measurements with cross-circularly polarized pulses, (**a,b**), and co-circularly polarized pulses (**c,d**). These show that with the first two pulses having opposite circular polarizations the bound biexciton coherence can be excited. In contrast, with co-circular excitation the Pauli exclusion principle prevents the formation of biexcitons, allowing only excitation of the unbound two-exciton state. In the absence of any interactions pathways **c** and **d** would cancel each other out, however, asymmetry between the $X_K X_K - X_K$ and $X_K - g$ due to many body effects makes this cancellation imperfect, leading to the correlated two-exciton signal on the diagonal in Fig.2 (d).*